\documentclass[aps,prd,preprint,nofootinbib]{revtex4}
\usepackage{amsmath,amssymb,graphics,epsfig,multirow}
\newcommand{\newc}{\newcommand}
\newc{\bsym}{\boldsymbol}
\newc{\mrm}{\mathrm}
\newc{\ovl}{\overline}
\newc{\ovla}{\overleftarrow}
\newc{\ovra}{\overrightarrow}
\newc{\wtil}{\widetilde}
\newc{\eps}{\epsilon}
\newc{\tri}{\triangle}
\newc{\hc}{\dagger}
\newc{\PD}{\partial}
\newc{\ra}{\rightarrow}
\newc{\lra}{\leftrightarrow}
\newc{\CL}{{\mathcal{L}}}
\newc{\SL}{\!\!\!/}
\newc{\LH}{\hat{L}}
\newc{\RH}{\hat{R}}
\newc{\sWsq}{\sin^2\theta_\mathrm{W}}
\newc{\cWsq}{\cos^2\theta_\mathrm{W}}
\newc{\half}{\frac{1}{2}}

\begin{document}
\title{Exploring Non-Supersymmetric New Physics in Polarized M\o{}ller Scattering}
\author{We-Fu Chang}
\email{wfchang@phys.nthu.edu.tw}
\affiliation{Department of Physics, National Tsing Hua University, Hsin Chu 300, Taiwan}
\author{John N. Ng}
\email{misery@triumf.ca}
\affiliation{Theory Group, TRIUMF, 4004 Wesbrook Mall, Vancouver, B.C., Canada}
\author{Jackson M. S. Wu}
\email{jbnwu@itp.unibe.ch}
\affiliation{Center for Research and Education in Fundamental Physics, Institute for Theoretical Physics, University of Bern, Sidlerstrasse 5, 3012 Bern, Switzerland}

\date{\today}

\begin{abstract}
We study in an effective operator approach how the effects of new physics from various scenarios that contain an extra $Z'$ neutral gauge boson or doubly charged scalars, can affect and thus be tested by the precision polarized M\o{}ller scattering experiments. We give Wilson coefficients for various classes of generic models, and we deduce constraints on the parameter space of the relevant coupling constants or mixing angles from the results of the SLAC E158 experiment where applicable. We give also constraints projected from the upcoming 1~ppb JLAB experiment. In the scenario where the extra $Z'$ is light 
($M_{Z'} \ll M_W$), we obtain further constraints on the parameter space using the BNL $g-2$ result where it is useful. We find that the BNL deviation from the Standard Model cannot be attributed to a light extra $Z'$ neutral gauge boson.
\end{abstract}


\maketitle

\section{Introduction}
Polarized electron scattering has a long and illustrious history that began with the celebrated SLAC polarized $e\,D$ scattering~\cite{slac}, which was instrumental in putting the Sandard Model (SM) on a firm experimental ground. The latest experiment of this class is the polarized M\o{}ller scattering. As only the electrons are involved, theory calculations at quantum loop levels can be performed free of hadronic uncertainties making it an ideal process for precision SM testing; any deviation from expectations will also be a clear signal of new physics beyond the SM.

A while ago, an experiment of this type has been successfully completed~\cite{E158}, where parity-violating measurements have been made from the scattering of longitudinally polarized electrons -- either left-handed (LH) or right-handed (RH) -- on an unpolarized electron target. At a momentum transfer of
$Q^2 = 0.026$~GeV$^2$, the left-right asymmetry
\begin{equation}
A_{LR} \equiv \frac{d\sigma_L - d\sigma_R}{d\sigma_L + d\sigma_R} \,,
\end{equation}
has been measured to an accuracy of approximately 20~ppb, and the weak angle determined to be
$\sWsq = 0.2397 \pm 0.0010\,(\mathrm{stat.}) \pm 0.0008\,(\mathrm{syst.})$, with its running established at the $6\sigma$ level. In the SM, $A_{LR}$ arises from the interference between the electromagnetic and the weak neutral currents, and is given by
\begin{equation}
A_{LR}^{\mathrm{SM}}= \frac{G_\mu Q^2}{\sqrt{2}\pi\alpha}\frac{1-y}{1 + y^4 + (1-y)^4}
\left[1 - 4\sWsq^{\ovl{\,\mathrm{MS}}}\right] \,, \qquad
Q^2 = ys \,, \quad s = 2 m_e E_{\mathrm{beam}} \,,
\end{equation}
where $G_\mu = 1.16637(1) \times 10^{-5}$~GeV$^{-2}$, $\alpha^{-1} = 137.0356$, and terms involving the electron mass $m_e$ have been dropped when appropriate. The electroweak corrections matching this experimental accuracy and more is given in Ref.~\cite{CM}. The value of the weak angle at $Q^2 = 0$ is found to be $\sWsq^{\ovl{\,\mathrm {MS}}}= 0.23867 \pm 0.00016$ in an updated analysis~\cite{EM}. We use this value in all our numerical calculations below.

Recently, a 12~GeV energy upgrade to the intense electron facility at JLAB has been approved. This raises the possibility of measuring $A_{LR}$ to an accuracy of 1~ppb. Since the theoretical uncertainty can in principle be calculated to this accuracy, a successful measurement will be sensitive to the new physics at the several TeV scale, which is also the physics landscape to be explored by the LHC.

Also recently, an anomalous abundance of positron flux in cosmic radiation has been reported by the PAMELA
collaboration~\cite{PAMELA}. This confirms the excess previously observed by the HEAT~\cite{HEAT} and AMS~\cite{AMS} experiments. The ATIC~\cite{ATIC} and PPB-BETS~\cite{PPB} balloon experiments have also recorded similar excess. One possible origin is from a hidden sector and low scale dark matter annihilation~\cite{AW}. The hidden sector scale is $\mathcal{O}$(GeV), and so is at the opposite end of the new heavy physics. The characteristic of this class of model is that possible couplings to SM matter fields must be extremely weak. A very simple model was constructed previously to study this possibility~\cite{CNW}, and M\o{}ller scattering was found to give very stringent constraints.

Motivated by all these encouraging developments, we examine in detail how an improved $A_{LR}$ measurement
could shed light on new physics. We focus here on classes of non-superymmetric new physics and how they might manifest themselves in precision polarized M\o{}ller scattering. It is well known that $A_{LR}$ can be used to probe the existence of doubly charged scalars and extra $Z'$ bosons. The doubly charged scalars are interesting in their own right, and they also arise in Type II seesaw models for neutrino masses~\cite{numasses}. For extra $Z'$ bosons, the literature is particularly rich with well motivated models. These include models of grand unified theory (GUT) based on the $E_6$ or $SO(10)$ gauge group~\cite{GUT}, hidden sector $Z'$ models~\cite{CNW,St}, and left-right symmetric~\cite{LR} and warped extra-dimensional Randall-Sundrum (RS)~\cite{RS} models based on the gauge group
$SU(2)_L \times SU(2)_R \times U(1)_{B-L}$. The effects of a new particle such as the doubly charged scalar of the extra $Z'$ boson can be encoded in an effective operator approach, which for $A_{LR}$ involve only two dimension six operators:
\begin{equation}
\mathcal{O}_1 = \bar{e} \gamma^\mu \LH e \, \bar{e} \gamma_\mu \LH e \,, \quad
\mathcal{O}_2 = \bar{e} \gamma^\mu \RH e \, \bar{e} \gamma_\mu \RH e \,, \qquad
\LH = \half(1-\gamma_5) \,, \quad \RH = \half(1+\gamma_5) \,.
\end{equation}
In the next section, we describe this effective operator approach. 

The rest of the paper is organized as follows. In Sec.~\ref{sec:NPMs}, we discuss the classes of new physics models. Besides updating relevant known results, we also present a new calculation of the contributions in the Minimal Custodial RS (MCRS) model with bulk symmetry $SU(2)_L \times SU(2)_R \times U(1)_{B-L}$, and we give comparisons with the usual four dimensional left-right symmetric models. Sec.~\ref{sec:Conc} contains our conclusions. The details of the MCRS model and the specifics of the fermion localization are collected into Appendix.~\ref{app:MCRS}.

\section{Effective Operators}
\subsection{High Scale New Physics}
If the new physics is at a scale higher than the Fermi scale $v \simeq 247$~GeV, an effective operator approach is powerful and avoids much model dependence. Denoting the scale of new physics by $\Lambda$ and assuming that the SM gauge group holds between $\Lambda$ and $v$, the effective Lagrangian with dimension six four-lepton operators added is given by
\begin{align}\label{eq:OVI}
\CL &= \CL_{SM} + \CL_6 \,, \notag \\
-2\CL_6 &= \frac{c_{LL}}{\Lambda^2}(\bar{L}_a \gamma^\mu L_a)(\bar{L}_b \gamma_\mu L_b) +\frac{c_{LR}}{\Lambda^2}(\bar{L}_a \gamma^\mu L_a)(\ovl{e}_R \gamma_\mu e_R)
+\frac{c_{RR}}{\Lambda^2}(\ovl{e}_R \gamma^\mu e_R)(\ovl{e}_R \gamma_\mu e_R) \notag \\
&\quad +\frac{d_{LL}}{\Lambda^2}(\bar{L}_a \gamma^\mu L_b)(\bar{L}_b \gamma_\mu L_a) + h.c. \,,
\end{align}
where $a,b$ are $SU(2)$ indices, and $L=(\nu,e)^T_L$. The subscripts $L$ and $R$ denote the usual chiral projection. We have included only terms relevant for the calculation of $A_{LR}$, i.e. the first generation leptons. Note that for scattering of a polarized beam on a polarized target, there is also a scalar operator that contributes:
\begin{equation}
\CL_s= \frac{c_S}{2\Lambda^2}\ovl{L}e_R\,\ovl{e}_R L + h.c.
\end{equation}
A full analysis of dimension six leptonic operators can be found in Ref.~\cite{CNL}.

The Wilson coefficients $c$ and $d$ are given by specific models at $\Lambda$. One has to do a renormalization group (RG) running analysis to determine their values at the relevant low energy scale where the experiments are performed. The RG equations for these coefficients can be found in Ref.~\cite{CNL}.
Numerically, we find that the RG running effects are not significant. They give about a $4\%$ effect running between $\Lambda$ (typically a few TeV) and $v$, and hence can be neglected in the first approximation. Below $v$ the running is governed by the well known $\beta_{QED}$ funtions. Including them leads to a change of about $10\%$ in the values of the coefficients between $\Lambda$ and $\sqrt{s} \simeq 0.1$ GeV. We neglect these small RG running effects below which would not impact our analysis given the current level of precision in the experimental inputs and the exploratory nature of our study.

\begin{figure}[htbp]
\begin{center}
\includegraphics{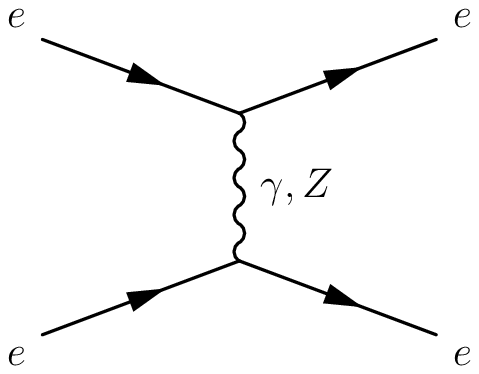}
\hspace{0.5in}
\includegraphics{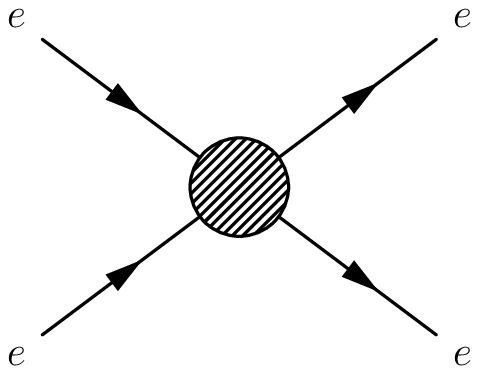}
\end{center}
\caption{\label{fig:Feyn} Feynman diagrams contributing to $A_{LR}$. The shaded blob contains all new physics effects.}
\end{figure}

The asymmetry $A_{LR}$ is found by calculating the diagrams depicted in Fig.~\ref{fig:Feyn}
The shaded blob schematically includes all new physics effects such as $Z$-$Z^\prime$ mixing and virtual exchanges of new particles. For $\Lambda^2 \gg s \gg m_e^2$ valid for the 12~GeV experiment, one has
\begin{align}\label{eq:ALR}
A_{LR} &= A_{LR}^\mathrm{SM} + \delta A_{LR} \notag \\
&= \frac{4G_\mu s}{\sqrt{2}\pi\alpha}\frac{y(1-y)}{1+y^4 +(1-y)^4}
\left[\left(\frac{1}{4}-\sWsq^{\ovl{\,\mathrm{MS}}}\right)
+\frac{c^{\prime}_{LL}-c_{RR}}{4\sqrt{2}G_\mu\Lambda^2}\right] \,,
\end{align}
where $y = -\frac{t}{s}$, $c^{\prime}_{LL} = c_{LL} + d_{LL}$, and $\delta A_{LR}$ denotes the deviation of the asymmetry from the SM prediction arising from new physics. The one-loop SM radiative corrections are encoded in the precise value of $\sWsq$ evaluated in the $\ovl{\mathrm{MS}}$ scheme. The effective operators, $\mathcal{O}_{1,2}$, can arise from numerous new physics models, and those coming from the tree-level exchange of virtual particles are the most important. We thus examine models with at least an extra neutral gauge boson, $Z'$, or doubly charged scalars below in Sec.~\ref{sec:NPMs}, where we give Wilson coefficients in various models. Of course there could be extra charged gauged bosons that also contribute in specific models, but these typically come in at the loop level and so are less important.

We remark here that the new physics effects in M\o{}ller scattering can also come in through the $Zee$ coupling being modified. This is particularly evident in cases when the heavy states of the new physics also have very small couplings to electrons so that the virtual exchange process is suppressed both by the propagator effect and the small coupling. Example of this can be found in the extra hidden sector/shadow $U(1)$ and RS models which we discuss below in Sec.~\ref{subsec:ShZ} and~\ref{subsec:RSM}.

\subsection{Low Scale New Phyiscs}
For the class of models where new physics arise at a scale $\Lambda \lesssim v$, two effects are at work. The first is due to the exchange of the new low mass particle. The propagator effect has to be compensated by its small couplings to the electrons of different chiralities. The second is due to the modification of the SM $Z$ coupling to the electrons. Well known examples are models with extra $Z'$ that can mix with the $Z$. Details are given below in Sec.~\ref{sec:NPMs}.

\section{\label{sec:NPMs} Model Considerations}
In this section, we discuss how new physics can be probed by polarized M\o{}ller scattering in various non-supersymmetric extensions of the SM. We begin with models that contain extra neutral $Z'$ bosons. A comprehensive review on the physics of $Z'$ can be found in Ref.~\cite{Zprime}.

Irrespective of the ultimate origin of the extra neutral gauge boson, its interactions can be described by the gauge group, $SU(2)_L \times U(1)_Y \times U(1)'$, at a scale not too high compare to $v$~\cite{yale}. Denoting the extra $U(1)'$ gauge field by $X_\mu$, its interaction with the SM matter content is given by
\begin{equation}\label{eq:LagX}
\CL^\prime = -\frac{1}{4}X_{\mu\nu}X^{\mu\nu} - \frac{\eps}{2}B_{\mu\nu}X^{\mu\nu}
+ g^\prime z_f \bar{f} \gamma^\mu f X_\mu - \frac{1}{2} M_X^2 X_\mu X^\mu + |D_\mu H|^2 \,,
\end{equation}
with the covariant derivative given by
\begin{equation}
D_\mu = \partial_\mu - i g_L\,T^a W^a_\mu - i\frac{g_Y}{2} B_\mu - i\frac{g'}{2} z_H X_\mu \,,
\end{equation}
where $g_L$, $g_Y$, and $g'$ are the gauge couplings of the $SU(2)_L$, $U(1)_Y$, and $U(1)'$ respectively, $f$ a species label that runs over all SM fermions, and $z_f$ ($z_H$) the SM fermion (Higgs) charge under the $U(1)'$. Other notations are standard. The paramter $\epsilon$ characterizes the kinetic mixing between $B_\mu$, the hypercharge gauge field, and $X_\mu$, and is expected in general to be of order $10^{-4}$ to $10^{-2}$~\cite{CNW,DKR}.

The kinetic energy terms for the gauge bosons can be recast into canonical form through a $GL(2)$ transformation
\begin{equation}\label{eq:KinMix}
\begin{pmatrix}
X \\
B
\end{pmatrix}
=
\begin{pmatrix}
 c_\eps & 0 \\
-s_\eps & 1
\end{pmatrix}
\begin{pmatrix}
X' \\
B'
\end{pmatrix} \,, \qquad
s_\eps = \frac{\eps}{\sqrt{1-\eps^2}} \,, \quad
c_\eps = \frac{1}{\sqrt{1-\eps^2}} \,.
\end{equation}
In most cases involving high scale new physics, the kinetic mixing is negligible. However, this is not true in general for low scale cases. The mass term in Eq.~\eqref{eq:LagX} can arise from a hidden sector scalar interactions that breaks the $U(1)^\prime$ gauge symmetry. The details are not important for this study. After the spontaneous breaking of the electroweak symmetry, this mass term will lead to a mass mixing among $W_3$, $B$, and $X$ fields which depend on $z_H$.

For models with $z_H = 0$ but a non-vanishing kinetic mixing between the neutral gauge bosons, the transformation between the weak and mass basis is given by
\begin{equation}
\begin{pmatrix}
B' \\
W_3 \\
X'
\end{pmatrix}
=
\begin{pmatrix}
c_W & -s_W & 0 \\
s_W &  c_W & 0 \\
0   &  0   & 1
\end{pmatrix}
\begin{pmatrix}
1 & 0      &  0 \\
0 & c_\eta & -s_\eta \\
0 & s_\eta &  c_\eta
\end{pmatrix}
\begin{pmatrix}
\gamma \\
Z \\
Z'
\end{pmatrix} \,,
\end{equation}
where $s_W$ ($c_W$) denotes $\sin\theta_W$ ($\cos\theta_W$), and similarly for the angle $\eta$. The first rotation is the standard one that give rise to the would be SM $Z$, which we label as $Z_0$, while the second diagonalizes the mixing between $Z_0$ and $X'$ that produces the two neutral gauge eigenstates, $Z$ and $Z'$. This mixing angle is given by
\begin{equation}\label{eq:t2eta}
\tan{2\eta} = \frac{2s_W s_\eps}{x^2 + s_W^2 s_\eps^2 - 1} \,, \qquad x = \frac{c_W M_X}{M_W} \,,
\end{equation}
and the masses of $Z$ and $Z'$ are given by
\begin{align}\label{eq:Zmasses}
M^2_\mp &= \frac{M_W^2}{2c_W^2}\left\{x^2 + 1 + s_W^2 s_\eps^2
\mp \sqrt{(x^2 - 1 + s_W^2 s_\eps^2)^2 + 4 s_W^2 s_\eps^2}\right\} \,.
\end{align}
where $M_-$ ($M_+$) denotes the lighter (heavier) of the two states. For high scale new physics, 
$M_W \ll M_X$. Since $s_\eps \lesssim 10^{-2}$, the mixing angle $\eta$ is very small as $\eta < \eps$ when $s_\eps$ is small and $x > 1$. For $M_W \gg M_X$, $\eps$ and $\eta$ can be of the same order. The relative sign between them depends on the relative magnitude between $x^2$ and $s_\eps^2$.

Models in which $\eps$ is vanishing but $z_H$ is not are more typical of high scale new physics scenarios. Since there is now no kinetic mixing, the $Z_0$ mixes with the $X$. The mass matrix of this system can be written as
\begin{equation}
M^2_{Z_0-X} = 
\begin{pmatrix}
M_{Z_0}^2 & \Delta^2 \\
\Delta^2  & M_X^2
\end{pmatrix} \,,
\end{equation}
with the mixing angle and the masses of the resulting $Z$ and $Z'$ given by~\cite{Zprime,yale}
\begin{equation}\label{eq:t2t}
\tan{2\theta} = \frac{2\Delta^2}{M_{Z_0}^2-M_X^2} \,,
\end{equation}
and
\begin{equation}\label{eq:Zmasses2}
M^2_{Z,\,Z'} = \frac{1}{2}\left[M_{Z_0}^2+M_X^2\mp\sqrt{(M_{Z_0}^2-M_X^2)^2+4\Delta^4}\right] \,.
\end{equation}
Note that both $\Delta$ and $M_X$ are model dependent. For example, if the SM Higgs is charged under both $U(1)$ factors, $\Delta^2 = -\frac{1}{8} z_H v^2 g^\prime \sqrt{g^2 + g_Y^2}$ and 
$M_X^2 = \frac{1}{4} v^2 g^{\prime 2} z_H^2 + M_0^2$ where $M_0$ is the mass from higher scale.

In either case, the Wilson coefficients can now be calculate from $\epsilon$, $g'$, $z_f$, $z_H$, and $\Lambda$ in a given model. For quick reference, we summarize here the results from below the Wilson coefficients for high scale new physics in Table~\ref{Tb:NPsumm}.
\begin{table}[htbp]
\begin{ruledtabular}
\begin{tabular}{ccccc}
New Physics & $c'_{LL}$ & $c_{RR}$  \\
\hline
$E_6$ GUT & $\frac{3\pi\alpha}{2c_W^2}\!\left(c_\beta+\sqrt{\frac{5}{27}}s_\beta\right)^2$ &
$\frac{3\pi\alpha}{2c_W^2}\!\left(\frac{1}{3}c_\beta-\sqrt{\frac{5}{27}}s_\beta\right)^2$ \\
$U(1)_R \times U(1)_{B-L}$ & $\frac{1}{4}\frac{g_Y^4}{g_R^2-g_Y^2}$ & $\frac{1}{4}\frac{(2g_Y^2-g_R^2)^2}{g_R^2 -g_Y^2}$\\
$U(1)_{B-L}$ & $g^{\prime 2}$ & $g^{\prime 2}$\\
$U(1)_X$ & $g^{\prime 2}z_L^2$ & $g^{\prime 2}z_e^2$ \\
$P^{\pm\pm}$ & 0 & $-|y_{ee}|^2/2$ \\
\end{tabular}
\end{ruledtabular}
\caption{\label{Tb:NPsumm} Summary of Wilson coefficients $c'_{LL}$ and $c_{RR}$ for various generic high scale new physics models containing an extra $Z'$ neutral gauge boson or doubly charged scalars ($P^{\pm\pm}$).}
\end{table}

\subsection{GUT Models}
The appearance of extra $U(1)$ gauge groups is exemplified by the $E_6$ GUT model. One way they can arise is through the symmetry breaking chain
\begin{equation}
E_6 \ra SO(10) \times U(1)_\psi \ra SU(5) \times U(1)_\chi \times U(1)_\psi \,.
\end{equation}
The $U(1)$ factors here are not anomalous. There are also no kinetic mixing, and the mass mixings between the gauge bosons of the extra $U(1)$ factors and the SM $Z$ are phenomenologically found to be negligible. The $X$ boson above is now a linear combination of the $\psi$ and $\chi$ gauge bosons
\begin{equation}\label{eq:Xbeta}
X^\mu(\beta) = B^\mu_\psi\sin{\beta} + B^\mu_\chi\cos{\beta} \,,
\end{equation}
where $B^\mu_{\psi,\,\chi}$ are gauge fields of the $U(1)_{\psi,\,\chi}$ respectively. The $X^\mu$ is taken to be the lighter of the two neutral linear combinations. From the fermion charges given in Ref.~\cite{Zprime,Mar}, we find~\footnote{As can be seen from Eq.~\eqref{eq:t2t}, with $\Delta\sim\mathcal{O}(100)$~GeV and $M_X\gtrsim\mathcal{O}(1)$~TeV expected, the $Z_0-X$ mass mixing is small with $\sin\theta\lesssim\mathcal{O}(0.01)$. We thus follow common practice and neglect the small corrections to $A_{LR}$ due to the mass mixing and the concurrent modification in the gauge-fermion couplings, which are proportional to $(\sin\theta)^2$.}
\begin{equation}
c_{LL}' - c_{RR} = 
\frac{4\pi\alpha}{3c_W^2}c_\beta\left(c_\beta + \sqrt{\frac{5}{3}}s_\beta\right) \,,
\end{equation}
where $c_\beta$ ($s_\beta$) denotes $\cos\beta$ ($\sin\beta$).

\begin{figure}[htbp]
\centering
\includegraphics[width=3.8in]{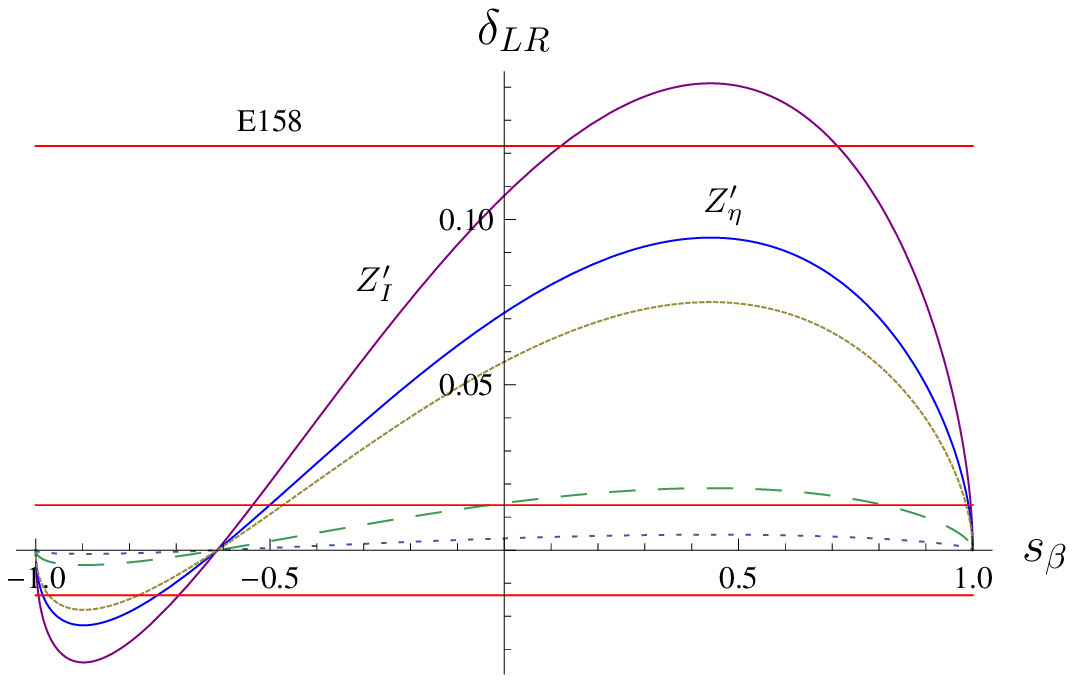}
\caption{\label{fig:GUTp} The ratio, $\delta_{LR}$, as a function of $s_\beta$. The labelled solid lines are the SLAC E158 upper limit, $\delta_{LR} = 0.122$, and cases for CDF lower mass limits on $Z_I'$ and $Z_\eta'$, $\Lambda = 729,\,891$~GeV. The other solid, dashed, and dotted lines denote cases where $\Lambda = 1,\,2,\,4$~TeV respectively. The narrow band marks the projected JLAB constriant, 
$|\delta_{LR}| < 0.0136$. All limits are given at $95\%$ CL.}
\end{figure}

Defining $\delta_{LR} \equiv (A_{LR} - A_{LR}^{SM})/A_{LR}^{SM} = \delta A_{LR}/A_{LR}^{SM}$, we plot in Fig.~\ref{fig:GUTp} the resulting $\delta_{LR}$ due to GUT new physics.
We plot also the constraints from the Tevatron new physics searches, the SLAC E158 experiment, and that projected from the the JLAB 1 ppb measurement. We use the Tevatron limits on the $Z_I'$ and $Z_\eta'$ masses to illustrate the current collider constraints on the new physics scale $\Lambda$ for the GUT extra neutral gauge bosons~\footnote{In our notation, $Z_I'$ and $Z_\eta'$ are just $X$ with $\tan\beta = \sqrt{3/5}$ and $-\sqrt{5/3}$ respectively. See Eq.~\eqref{eq:Xbeta}.}. The $95\%$ confidence level (CL) lower mass limit for $Z_I'$ ($Z_\eta'$) is
$M_{Z_I'} > 729$~GeV ($M_{Z_\eta'} > 891$~GeV), representing the lowest (highest) exclusion limit in the CDF search~\cite{CDF}. The SLAC E158 measurement of the left-right asymmetry~\cite{E158} gives
\begin{equation}
A_{LR} = 131 \pm 14\,(\mrm{stat}) \pm 10\,(\mrm{syst})\,\mrm{ppb} \,.
\end{equation}
Taking $A_{LR}^{SM} = 1.47 \times 10^{-7}$ for $Q^2 = 0.026$~GeV$^2$ and $y = 0.6$, one can infer from this that $-0.337 < \delta_{LR} < 0.122$ at $95\%$ CL~\footnote{We have followed the more common practice here in taking the $1\sigma$ total error as the statistical and systematic uncertainties added in quadrature. The $95\%$ CL corresponds to an error interval of $\pm 1.96\sigma$.}. Lastly, as a benchmark assume that the total $1\sigma$ error of the new JLAB measurement of $A_{LR}$ to be 1~ppb with a central value given by the SM value, $A_{LR}^{SM}$, one expects in such a case $|\delta_{LR}| < 0.0136$ at $95\%$ CL. We see from Fig.~\ref{fig:GUTp} that current experiments do not constrain TeV scale GUT new physics, but the upcoming 1~ppb JLAB experiment can be expected to do so.

\subsection{\label{subsec:LRSM} Left-Right Symmetric Models}
\begin{figure}[htbp]
\centering
\includegraphics[width=3.6in]{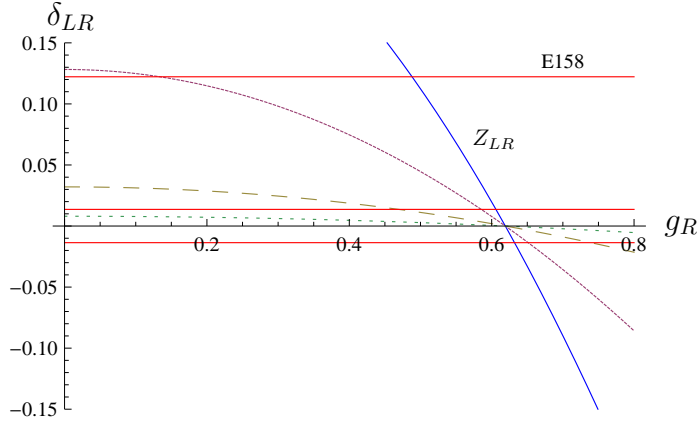}
\caption{\label{fig:LRSp} The ratio, $\delta_{LR}$, as a function of $g_R$. The labelled solid lines are the SLAC E158 upper limit, $\delta_{LR} = 0.122$, and the case for the $Z_{LR}$ lower mass limit, $\Lambda = 630$~GeV. The other solid, dashed, and dotted lines denote cases where $\Lambda = 1,\,2,\,4$~TeV respectively. The narrow band marks the projected JLAB constraint, $|\delta_{LR}| < 0.0136$. All limits are given at $95\%$ CL.}
\end{figure}

For left-right symmetric models, the gauge group is taken to be $SU(2)_L \times SU(2)_R \times U(1)_{B-L}$. The $SU(2)_R$ is broken down to a $U(1)_R$ by a triplet Higgs field at a scale greater than a few TeV, leaving $SU(2)_L \times U(1)_R \times U(1)_{B-L}$. In this case, the $U(1)_R \times U(1)_{B-L}$ is equivalent to the $U(1)_Y \times U(1)^\prime$~\cite{yale}. The charges $z_f$ and $z_H$ can be determined by enforcing anomaly cancellation conditions (see below). We find
\begin{equation}
c_{LL}' - c_{RR} = \frac{1}{4}\left(3g_{Y}^2-g_{R}^2 \right)\,,
\end{equation}
where $g_R$ is the gauge coupling of $SU(2)_R$. Interestingly, note that even if $g_R = g_L$, $\delta_{LR}$ will not vanish. We plot in Fig.~\ref{fig:LRSp} the resulting $\delta_{LR}$ for left-right symmetric models, as well as the relevant SLAC E158 constraint and that projected from the JLAB 1~ppb measurement. As a comparion, we plot also the curve that arises from the $Z_{LR}$ $95\%$ CL lower mass limit, $M_{Z_{LR}} > 630$~GeV, obtained from the combination of electroweak data, direct Tevatron and indirect LEP II searches~\cite{Zprime}. We see that current experiments do offer, if only barely, some constraints on the TeV scale new physics here compared to the GUT case. But for useful ones, the precision of the upcoming JLAB experiment is still much needed.

\subsection{$SU(2)_L \times U(1)_Y \times U(1)_X$ Models}
There are interesting bottom up models where the extra $U(1)$ factors are not immediately related to the remnant of some broken down non-abelian groups. The prime example is the
$SU(2)_L \times U(1)_Y \times U(1)_{B-L}$ model. The assumption is that the SM is charged under $U(1)_X$, and since this group is known to be anomalous, one has to demand anomaly cancellations for chiral fermions under the full $SU(3)_C \times SU(2)_L \times U(1)_Y \times U(1)_X$. As neutrinos are massive, we introduce one sterile neutrino, $N_R$, per family and denote its charge by $z_N$. Assuming these charges are family independent, we have seven independent charges $z_f$, $z_N$, and $z_H$. The anomaly conditions arise from
\begin{equation}
[SU(3)]^2 U(1)_X \,, \quad [SU(2)]^2 U(1)_X \,, \quad [U(1)_Y]^2 U(1)_X \,, \quad [U(1)_X]^3 \,.
\end{equation}
The mix gravitational and $U(1)_X$ anomaly give the same condition as $[U(1)_X]^3$ and is redundant. The Yukawa terms that give quark and lepton masses yielded one additional condition. Thus there are two independent charges, which we choose to be $z_L$ and $z_e$. The other charges are
\begin{equation}
z_Q = -\frac{1}{3}z_L \,, \quad z_u = \frac{2}{3}z_L-z_e \,, \quad z_d = -\frac{4}{3}z_L+z_e \,, \quad
z_N = 2z_L - z_e \,, \quad z_H = z_L - z_e \,.
\end{equation}

\begin{figure}[htbp]
\centering
\includegraphics[width=3.6in]{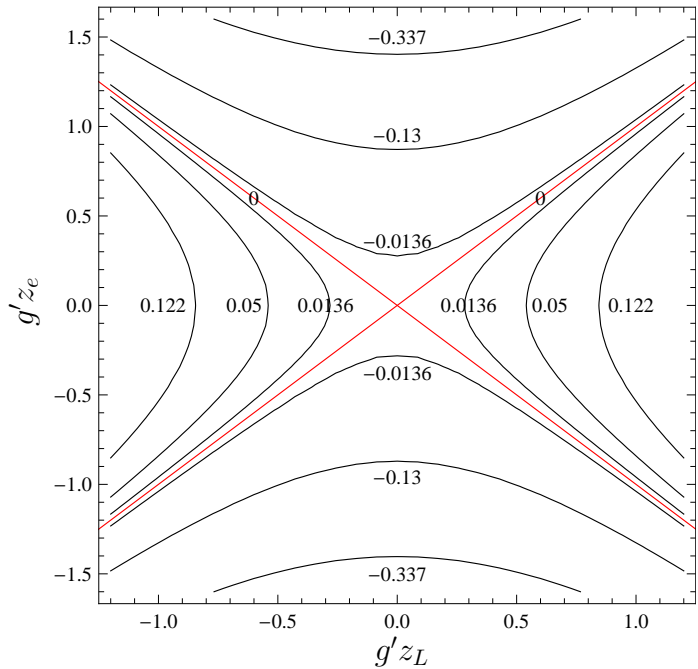}
\caption{\label{fig:Xc} Contours of constant $\delta_{LR}$ for $\Lambda = 1$~TeV. Note the contours for the SLAC E158 lower and upper limits, $\delta_{LR} = -0.337$ and $\delta_{LR} = 0.122$, and that for the  projected JLAB limits, $\delta_{LR} = \pm 0.0136$. All limits are given at $95\%$ CL.}
\end{figure}

For the interesting case where $X = B-L$, we have the familiar lepton and baryon number assignments, and
$z_H = 0$. More importantly, if the extra $Z'$ arises directly from the breaking of a gauged $U(1)_{B-L}$, M\o{}ller scattering will yield only the SM result, since in that case $z_L = z_e$; this leads immediately to $c'_{LL} - c_{RR} = 0$. In general, $z_L \neq z_e$, and we find
\begin{equation}
c_{LL}' - c_{RR} = g^{\prime 2}(z_L^2 - z_e^2) \,,
\end{equation}
where $g'$ is the gauge coupling constant of the $U(1)_X$. Note that in this analysis the masses of the sterile neutrinos do not come into play. Hence the details of the mechanism for generating active neutrinos masses will not change our results. We plot in Fig.~\ref{fig:Xc} contours of constant $\delta_{LR}$ as a function of $g' z_L$ and $g' z_e$ for $\Lambda = 1$~TeV. For other values of $\Lambda$, the contour plot is simply rescaled by a factor of $(\Lambda/1\,\mrm{TeV})$ as can be seen from Eq.~\eqref{eq:ALR}.

\subsection{\label{subsec:ShZ} Hidden Sector/Shadow $Z'$ Models}
Another class of models with an extra $U(1)$ factor, which has received renewed attention recently due to the observation of excess cosmic positron flux, is one such that there are no direct couplings of the extra gauge field to SM fermions or Higgs fields. The extra $U(1)_{hid}$ is a gauge symmetry of a hidden sector with its own scalar sector. The hidden sector and the SM communicate through hidden scalars couplings with the SM Higgs~\cite{CNW} and kinetic mixing~\cite{KW}. In this class of models $z_H = 0$, and the fermion couplings to $Z$ and $Z'$ are given by~\cite{CNW}
\begin{align}
\label{eq:zff}
i\gamma^\mu\frac{g_L}{c_W} &\left\{
\Big[c_\eta Q^L_Z(f) - \half s_\eta s_W s_\eps Y_f^L\Big]\LH + (L \lra R)\right\}\quad
(Z^\mu\bar{f}f) \,, \\
\label{eq:zprimeff}
-i\gamma^\mu\frac{g_L}{c_W} &\left\{
\Big[s_\eta Q^L_Z(f) + \half c_\eta s_W s_\eps Y_f^L\Big]\LH + (L \lra R)\right\}\quad
(Z^{\prime\mu}\bar{f}f) \,,
\end{align}
where $Q^{L,\,R}_Z(f) = T^3(f_{L,\,R}) - s_W^2 Q_f$ with $Q_f = T^3_f + Y_f/2$. We see that the neutral current couplings are not only rotated as indicated by the $c_\eta$ factor, but they also contain an extra piece proportional to the fermion hypercharge due to the $U(1)$-$U(1)_X$ mixing. Hence we need to re-examine the electroweak precision data using the full couplings as well as taking into account the effects due to virtual $Z'$ exchanges.

As mentioned above, there are two interesting limits with very different phenomenologies depending on whether $M_X$ is greater or less than $M_W$, which correspond to high or low scale new physics.

\subsubsection{$M_X \gg M_W$}
From Eq.~\eqref{eq:t2eta} we can see that $\eta \ll \eps$ and the mixing can be ignored. The most stringent constraints come from electroweak precision measurements at the $Z$-pole. As for M\o{}ller scattering, the exchange of the $Z'$ gives negligible effect because its large mass and very small coupling to the electrons (see Eq.~\eqref{eq:zprimeff}). The dominant effect comes from the modified $Z e\bar{e}$ coupling, and we find
\begin{align}\label{eq:zee}
\delta_{LR}\equiv\frac{\delta A_{LR}}{A_{LR}^\mrm{SM}} &\simeq
-s_\eta^2 - 3 \frac{s_\eta^2 s_W^2 s_\eps^2}{1 - 4s_W^2}
-2c_\eta s_\eta s_W s_\eps \frac{2s_W^2 + 1}{1 - 4s_W^2} \notag \\
&= -\big(s_\eta^2 + 15.80 s_\eta^2 s_\eps^2 + 31.85 c_\eta s_\eta s_\eps\big) \,,
\end{align}
where the Thomson limit value of the weak angle, $s_W^2 = 0.23867$, is used. Unless otherwise specified, we will use this value of $s_W^2$ for all numerical presentations below. We point out here that this result is independent of $M'$. Also if $s_\eta$ and $s_\eps$ have the same sign, $\delta_{LR}$ is negative.

\subsubsection{$M_X \ll M_W$}
\begin{figure}
\centering
\includegraphics[width=3.2in]{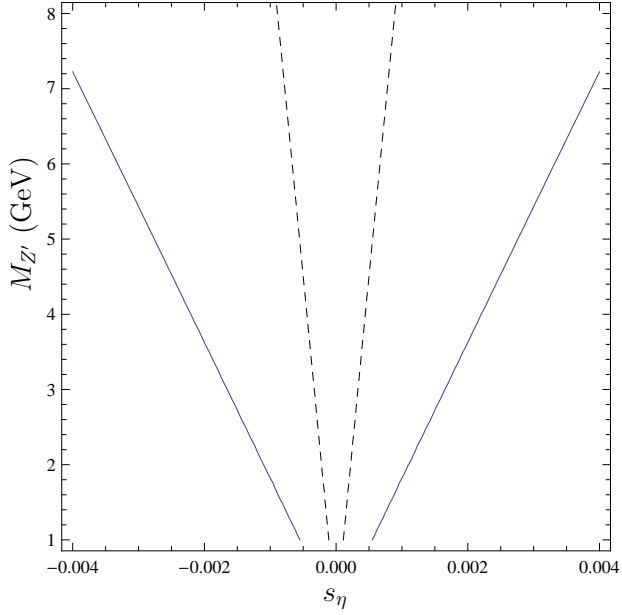}
\caption{\label{fig:SLAC} Constraint from the light $Z'$ contribution to $A_{LR}$. The contours are that for the SLAC E158 lower limit (solid), $\delta_{LR} = -0.337$, and the projected JLAB lower limit (dashed), $\delta_{LR} = -0.0136$. All limits are given at $95\%$ CL. The allowed region is in between the respective contours. No parameter space exists for $\delta A_{LR} > 0$.}
\end{figure}

In this case we have a light $Z'$ whose mass is essentially $M_X$ (see Eq.~\eqref{eq:Zmasses}), and the mixing angle $\eta$ can be large compare to $\eps$. Since the $Z'$ is light, $\delta_{LR}$ is dominated by the photon-$Z'$ interference term:
\begin{equation}
\delta_{LR}\approx\delta_{LR}^{Z'} \simeq
4\,\frac{(u^L_e)^2 - (u^R_e)^2}{1-4s_W^2}\frac{M_Z^2}{M_{Z'}^2} =
\left(s_\eta^2 - 15.80 c_\eta^2 s_\eps^2 + 31.85 c_\eta s_\eta s_\eps\right)\frac{M_Z^2}{M_{Z'}^2} \,, \end{equation}
where $u^{L,\,R}_e = -\big[s_\eta Q^{L,R}_Z(e) + \half c_\eta s_W s_\eps Y_e\big]$. From the SLAC E158 measurement of $A_{LR}$, we have at $95\%$ CL
\begin{equation}\label{eq:E158c}
s_\eta^2 - 15.80 c_\eta^2 s_\eps^2 + 31.85 c_\eta s_\eta s_\eps \in 
[-4.06,1.47] \times 10^{-5} \left(\frac{M_{Z'}}{1\,\mrm{GeV}}\right)^2 \,.
\end{equation}
If we use the projected JLAB limits from above, the constraint interval on the RHS is change to $[-0.16,0.16]$ in the same unit. Now from Eqs.~\eqref{eq:t2eta} and~\eqref{eq:Zmasses}, $s_\eps$ can be written in terms of $M_{Z'}$ and $s_\eta$ as
\begin{equation}\label{eq:sZs}
s_\eps = -\frac{s_\eta}{s_W c_\eta}\left(1 - \frac{c_W^2 M_{Z'}^2}{M_W^2}\right) \,.
\end{equation}
Using this, we plot in Figs~\ref{fig:SLAC} the allowed $(s_\eta,M_{Z'})$ region as given by Eq.~\eqref{eq:E158c}.
Note that for the RHS of Eq.~\eqref{eq:E158c} greater than zero, there is no real solution, and so no allowed parameter space.

The mixing parameters $\eta$ and $\eps$ are also constrained by low energy observables such as the anomalous magnetic moment of the muon $a_\mu$. The new physics contribution from a light extra $Z'$ to $a_\mu$ is found to be~\cite{CNW}
\begin{align}
\delta a_\mu &= \frac{G_\mu m_\mu^2}{6\sqrt{2}\pi^2}\frac{M_Z^2}{M_{Z'}^2}
\big[s_\eta^2(4s_W^4-2s_W^2-1) - 2(3s_W^2-2)s_W s_\eta c_\eta s_\eps + s_W^2 c_\eta^2 s_\eps^2\big] 
\notag \\
&= -1.94 \times 10^{-9}\big(s_\eta^2 - 0.19 c_\eta^2 s_\eps^2 - 1.00 c_\eta s_\eta s_\eps\big)
\frac{M_Z^2}{M_{Z'}^2} \notag \\
&= -1.94 \times 10^{-9}s_\eta^2\left\{1 + \frac{M_W^2 - c_W^2 M_{Z'}^2}{M_W^4 s_W^2}
\Big[M_W^2(s_W - 0.19) + 0.19 M_{Z'}^2\Big]\right\}\frac{M_Z^2}{M_{Z'}^2} \,,
\end{align}
where we have used Eq.~\eqref{eq:sZs} in the last line. We see that for $M_{Z'} \ll M_W$, the curly bracket is positive definite. Thus a $\delta a_\mu$ arising from a light $Z'$ is always negative. This means that if confirmed, the current BNL result of a $3.4 \sigma$ deviation from SM expectation~\cite{BNL}:
\begin{equation}
a_\mu^\mrm{exp}-a_\mu^\mrm{SM} = (2.95 \pm 0.88) \times 10^{-9} \,,
\end{equation}
cannot be due to the new physics arising from the light $Z'$ alone.

\subsection{\label{subsec:RSM} The MCRS Model}
Recently, the warped five-dimensional (5D) RS model endowed with a bulk gauge group
$SU(3)_c \times SU(2)_L \times SU(2)_R \times U(1)_X$~\cite {ADMS} -- which we call the MCRS model -- has been actively studied as a framework for flavor physics~\cite{CNW08,CFW,Mainz}. The gauge hierarchy problem
is solved when the warp factor is taken to be $k \pi r_c \approx 37$~\cite{RS}. This sets the scale for the
new physics to $m_{KK} \approx 4$~TeV where the lowest Kaluza-Klein (KK) excitation first arises~\footnote{Note this is at the higher end of the LHC discovery limits.}. The bulk $SU(2)_R$ factor protects the $\rho$ parameter in the effective 4D theory from excessive corrections coming from KK excitations, and is broken by orbifold boundary conditions on the UV boundary (brane); this is a 5D generalization of the case studied in Sec.~\ref{subsec:LRSM}. A brief summary of the model is given in Appendix~\ref{app:MCRS}, which sets our notations.

In the MCRS model, $\delta_{LR}$ arise due to the interactions of KK excitations of the neutral gauge bosons and the charged leptons. Contributions come either from direct virtual exchanges of neutral gauge KK modes, or through the $Z e\bar{e}$ coupling been modified due to mixings of SM modes with KK modes. The latter involve both fermions and gauge bosons, and is similar to the shadow $Z'$ case discussed above. More details of the vertex correction due to KK mixings can be found in Appendix~\ref{app:MCRS}.

A generic feature of the extra dimensional models is that gauge-$f\bar{f}$ couplings are not flavor diagonal in the mass eigenbasis. To calculate the contributions to $\delta_{LR}$ in the mass eigenbasis requires the knowledge of both the LH and RH rotation matrices, which are determined by the mass matrix of the charged leptons and their localization in the extra dimension (see Ref.~\cite{CNW08}). In Table~\ref{Tb:RSALR}, we list the contributions pertinent to the case of polarized M\o{}ller scattering~\footnote{We do not displayed the tiny imaginary parts, which is $\lesssim\mathcal{O}(10^{-21})$, since the real part of the contribution dominates in the tree-level $A_{LR}$.}.
\begin{table}[htbp]
\begin{ruledtabular}
\begin{tabular}{cccccccc}
\multirow{2}{*}{Config.} 
& \multicolumn{3}{c}{Direct KK exchange} & \multicolumn{3}{c}{$Z e\bar{e}$ correction} & Total \\
& LL ($10^{-5}$) & RR ($10^{-6}$) & $\delta_{LR}^{dir}\,(10^{-4})$
& LL ($10^{-4}$) & RR ($10^{-5}$) & $\delta_{LR}^{Z e\bar{e}}\,(10^{-3})$ 
& $\delta_{LR}\,(10^{-3})$ \\
\hline
I   & 1.98574 & 1.51504 & 8.09458 & -1.07684 & -7.98263 & -2.45877 & -1.64931 \\
II  & 1.98956 & 1.51484 & 8.11154 & -1.07827 & -7.98167 & -2.47224 & -1.66109 \\
III & 1.99197 & 1.51050 & 8.12409 & -1.07920 & -7.96023 & -2.49935 & -1.68694 \\
IV  & 1.99188 & 1.51065 & 8.12365 & -1.07916 & -7.96172 & -2.49767 & -1.68530 \\
V   & 1.99049 & 1.51483 & 8.11566 & -1.07858 & -7.98162 & -2.47504 & -1.66347 \\
\end{tabular}
\end{ruledtabular}
\caption{\label{Tb:RSALR} New physics contributions to the polarized M\o{}ller scattering in the MCRS model for five typical charged lepton configurations that give rise to experimentally observed charged lepton masses, and are admissible under current LFV constraints.}
\end{table}
We give results for five typical charged lepton configurations that lead to charged lepton mass matrices compatible with the current lepton flavor violation (LFV) constraints, and give rise to experimentally observed charged lepton masses. Details of the charged lepton configurations and the particular realization of the admissible charged lepton mass matrices used to calculate the new physics contributions in Table~\ref{Tb:RSALR} are given in Appendix~\ref{app:LepM}. Note that the direct KK LL (RR) contribution corresponds to the Wilson coefficient $c_{LL}'$ ($c_{RR}$) scaled by a factor $(2\sqrt{2}G_F)^{-1}$.

As can be seen from Table~\ref{Tb:RSALR}, the new physics contribution to $A_{LR}$ from direct exchanges of virtual KK gauge modes are subdominant compare to that from the shift in the SM $Z e\bar{e}$ coupling due to KK mixing effects. This is because the gauge KK modes are heavy~\footnote{As required by electroweak precision measurements and very stringent quark sector flavor constraints, $m_{KK} \gtrsim 4$~TeV} and couplings of the electron to gauge KK modes are small. Note that the LL-type contribution is always greater that of the RR-type. This is a consequence of the fact that the LH charge is greater than the RH charge, i.e. $|Q_Z^L(e)| > |Q_Z^R(e)|$. Note also the new physics contributions are quite stable across the configurations, indicating their robustness with respect to the variations in the charged lepton mass matrix. The MCRS values of $\delta_{LR}$ at $m_{KK} \approx 4$~TeV are well within the projected JLAB limits of $|\delta_{LR}| < 0.0136$.

\subsection{Doubly Charged Scalars}
\begin{figure}[htbp]
\centering
\includegraphics[width=3.6in]{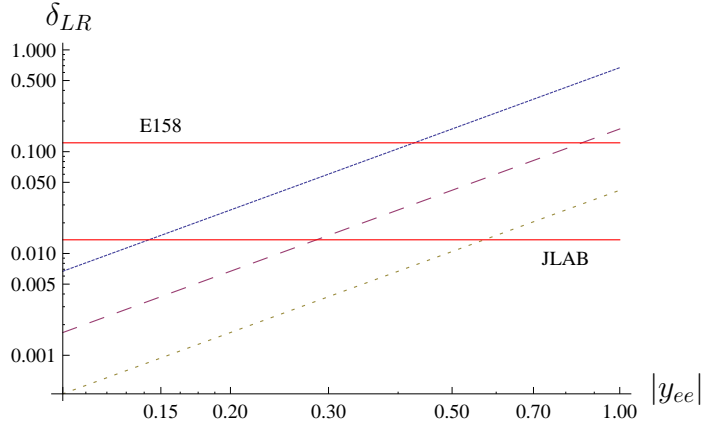}
\caption{\label{fig:DCp} The ratio, $\delta_{LR}$, as a function of $|y_{ee}|$. The solid, dashed, and dotted lines denote cases where $\Lambda = 1,\,2,\,4$~TeV respectively. The horizontal lines marks the SLAC E158 upper limit, $\delta_{LR} = 0.122$, and the projected JLAB upper limit, $\delta_{LR} = 0.0136$. All limits are given at $95\%$ CL.}
\end{figure}

Doubly charged scalars $P^{\pm\pm}$ are motivated by Type II seesaw neutrino mass generation mechanisms. They can be either $SU(2)_L$ singlets or triplets, and they carry two units of lepton number. The contribution of $P^{\pm\pm}$ to M\o{}ller scattering depends on only two parameters, viz. its mass, $M_P$, and its coupling to electrons, $y_{ee}$. For example, if there is only coupling to RH electrons, the effective contact interaction is
\begin{equation}
\frac{|y_{ee}|^2}{2M_P^2}(\ovl{e^c_R}e_R)(\bar{e}_R e^c_R) =
\frac{|y_{ee}|^2}{4M_P^2}(\bar{e}\gamma^\mu\RH e)(\bar{e}\gamma_\mu\RH e) \,,
\end{equation}
leading to
\begin{equation}
\delta_{LR} = \frac{|y_{ee}|^2}{2\sqrt{2}(1-4s_W^2)G_\mu M_P^2} \,.
\end{equation}
For models where the coupling is to LH electrons, $\delta_{LR}$ is similarly given with the RR Wilson coefficient changed to that of the LL type. The important difference is that $\delta_{LR}$ is negative here. We plot the $\delta_{LR}$ arising from doubly charged scalars coupling only to RH electrons in Fig.~\ref{fig:DCp}, as well as the relevant constraints from the SLAC E158 and the projected JLAB 1~ppb measurement.

\section{\label{sec:Conc} Conclusions}
We have examined in detail the implications of a very high precision polarized M\o{}ller scattering experiment on new physics beyond the SM, which is well known to be sensitive to extra $Z'$ and doubly charged scalars in the few TeV mass range. We have shown how such a measurement can be particularly useful if such a state is found at the LHC. If a doubly charged scalar were discovered at the LHC, a measurement of the $\delta_{LR}$ will determine its Yukawa couplings to electrons, and the sign of $\delta_{LR}$ will determine whether it couples to LH or RH electrons (negative for the former, positive for the latter). Such scalars also carry two units of lepton number, hence measurements of $\delta_{LR}$ will in turn have impact on the predicted rates of neutrinoless double beta decays of nuclei~\cite{numasses}, thus providing invaluable information on the origin of neutrino masses. Similarly, if an extra $Z'$ were discovered, measuring $\delta_{LR}$ will shed light on its origin. We have given detailed calculations for various classes of models including the $E_6$ GUT model, left-right symmetric model, and two examples of non-anomalous $U(1)$.

For high scale new physics, we found if the new particle is of the more exotic type such as a hidden/shadow $Z'$ or the pure KK $Z'$ in warped extra-dimension RS scenarios, $\delta_{LR}$ measures the change in the effective $Z e\bar{e}$ couplings as compare to the SM, and it is not sensitive to the mass of these new gauge bosons. In the scenario where the new physics is of a very low scale ($\ll v$), and a $Z'$ with a small mass exists that couples very weakly to SM fermions, a $\delta_{LR}$ measurement gives even more valuable information not available from high energy hadron collider experiments. We have shown that the SLAC E158 experiment can already tightly constrains the mixings of this type of $Z'$ with the SM $Z$. The smallness of the mixings allowed conforms with string-theoretic expectations, as well as phenomenological determinations~\cite{CNW}. On the other hand, the positive deviation from the SM expectation in the measured value of muon $g-2$ cannot be accommodated by such a light shadow $Z'$ since it gives a negative contribution. To the best of our knowledge there is no cleaner or better measurements on couplings of a hidden/shadow $Z'$ to electrons. These are valuable inputs to dark matter model building employing a hidden or shadow sector.

It is clear that an extra $Z'$ or doubly charged scalars with masses in the 1 to 2~TeV range can be found at the LHC relatively easily. We have shown that $\delta_{LR}$ can provide cross checks on these measurements as well as independent information on the couplings to electrons of different chiralities. If the masses
are $\gtrsim 4$ TeV, discovery at the LHC will take longer, and may have to await an intensity upgrade to the Super LHC (SLHC). Then it is even more important to do a high precision $\delta_{LR}$ measurement. The current reach of new physics of a 10 ppb measurement achieved at SLAC is around 1 to 2~TeV. Given the same couplings, a 1\% measurement will probe new physics at a scale $\Lambda \sim 3 - 6$~TeV, and so is nicely complementary to the SLHC. This is true in particular for the MCRS model, as the quark sector flavor bounds already constrained the masses of the lowest KK modes to be in at or above this range.

\begin{acknowledgments}
The research of W.F.C. is supported by the Taiwan NSC under Grant No. 96-2112-M-007-020-MY3.
The research of J.N.N. is partially supported by the Natural Science and Engineering Council of Canada. 
The research of J.M.S.Wu is supported in part by the Innovations und Kooperationsprojekt C-13 of the Schweizerische Universitaetskonferenz SUK/CRUS.
\end{acknowledgments}

\appendix

\section{\label{app:MCRS} RS model with $SU(2)_L \times SU(2)_R \times U(1)_{B-L}$ bulk symmetry}
We describe briefly in this appendix the basic set-up of the MCRS model to establish notations relevant for studying flavor physics. A more detailed description can be found in, e.g.~Ref.~\cite{ADMS}.

The MCRS mode is formulated on a slice of $AdS_5$ space specified by the metric
\begin{equation}\label{Eq:metric}
ds^2 = G_{AB}\,dx^A dx^B
= e^{-2\sigma(\phi)}\,\eta_{\mu\nu}dx^{\mu}dx^{\nu}-r_c^2 d\phi^2 \,,
\end{equation}
where $\sigma(\phi) = k r_c |\phi|$, $\eta_{\mu\nu} = \mathrm{diag}(1,-1,-1,-1)$, $k$ is the $AdS_5$ curvature, and $-\pi\leq\phi\leq\pi$. The theory is compactified on an $S_1/(Z_2 \times Z_2')$ orbifold, with $r_c$ the radius of the compactified fifth dimension, and the orbifold fixed points at $\phi= 0$ and $\phi=\pi$ correspond to the UV (Planck) and IR (TeV) branes respectively. To solve the hierarchy problem, one takes $k \pi r_c \approx 37$. The warped down scale is defined to be $\tilde{k} = k e^{-k\pi r_c}$. Note that $\tilde{k}$ sets the scale of the first KK gauge boson mass, $m^{(1)}_{gauge} \approx 2.45\tilde{k}$, which determines the scale of the new KK physics.

The MCRS model has a bulk gauge group $SU(3)_c \times SU(2)_L \times SU(2)_R \times U(1)_X$ under which the IR brane-localized Higgs field and transforms as $(1,2,2)_0$. The SM fermions are embedded into
$SU(2)_L \times SU(2)_R \times U(1)_X$ via the five-dimensional (5D) bulk Dirac spinors
\begin{align}\label{Eq:qrep}
Q_i &=
\begin{pmatrix}
u_{iL}\,[+,+] \\
d_{iL}\,[+,+]
\end{pmatrix} \,,\quad
U_i =
\begin{pmatrix}
u_{iR}\,[+,+] \\
\tilde{d}_{iR}\,[-,+]
\end{pmatrix} \,,\quad
D_i =
\begin{pmatrix}
\tilde{u}_{iR}\,[-,+] \\
d_{iR}\,[+,+]
\end{pmatrix} \,, \notag \\
L_i &=
\begin{pmatrix}
\nu_{iL}\,[+,+] \\
e_{iL}\,[+,+]
\end{pmatrix} \,,\quad
E_i =
\begin{pmatrix}
\tilde{\nu}_{iR}\,[-,+] \\
e_{iR}\,[-,+]
\end{pmatrix} \,, \qquad
i = 1,\,2,\,3 \,,
\end{align}
where $Q_i$ transforms as $(2,1)_{1/6}$, $U_i$ and $D_i$ as $(1,2)_{1/6}$, $L_i$ as $(2,1)_{-1/2}$, and $E_i$ as $(1,2)_{-1/2}$. The parity assignment $\pm$ denote the boundary conditions applied to the spinors on the $[\mathrm {UV}, \mathrm {IR}]$ brane, with $+$ ($-$) being the Neumann (Dirichlet) boundary conditions. Only fields with the [+,+] parity contain zero-modes that do not vanish on the brane. These survive in the low energy spectrum of the 4D effective theory, and are identified as the SM fields. Note that since we are only interested in the charged leptons in this work, we need only the RH charged leptons, and it is not necessary to have the doubling in the lepton sector as in the quark sector.

A given 5D bulk fermion field, $\Psi$, can be KK expanded as
\begin{equation}\label{Eq:PsiKK}
\Psi_{L,R}(x,\phi) = \frac{e^{3\sigma/2}}{\sqrt{r_c\pi}}
\sum_{n=0}^\infty\psi^{(n)}_{L,R}(x)f^n_{L,R}(\phi) \,,
\end{equation}
where subscripts $L$ and $R$ label the chirality, and the KK modes $f^n_{L,R}$ are normalized according to
\begin{equation}\label{Eq:fnorm}
\frac{1}{\pi}\int^\pi_{0}\!d\phi\,f^{n\star}_{L,R}(\phi)f^m_{L,R}(\phi) = \delta_{mn} \,.
\end{equation}
The KK-mode profiles are obtained from solving the equations of motion. For zero-modes, we have
\begin{equation}\label{eq:KK0}
f^0_{L,R}(\phi,c_{L,R}) =
\sqrt{\frac{k r_c\pi(1 \mp 2c_{L,R})}{e^{k r_c\pi(1 \mp 2c_{L,R})}-1}}
e^{(1/2 \mp c_{L,R})k r_c\phi} \,,
\end{equation}
where the c-parameter is determined by the bulk Dirac mass parameter, $m = c\,k$, and the upper (lower) sign applies to the LH (RH) label. Depending on the orbifold parity of the fermion, one of the chiralities is projected out.

After spontaneous symmetry breaking, the Yukawa interactions localized on the IR brane lead to mass terms for the fermions on the IR brane
\begin{equation}
S_\mrm{Yuk} = \int\!d^4x\,\frac{v_W}{k r_c\pi}\Big[
\ovl{Q}(x,\pi)\lambda^u_{5}U(x,\pi)+\ovl{Q}(x,\pi)\lambda^d_{5}D(x,\pi)+
\ovl{L}(x,\pi)\lambda^e_{5}E(x,\pi)\Big]+\mrm{h.\,c.} \,,
\end{equation}
where $v_W = 174$~GeV is the VEV acquired by the Higgs field, and $\lambda^{u,\,d,\,e}_5$ are (complex) dimensionless 5D Yukawa coupling matrices. For zero-modes, which are identified as the SM fermions, this gives the mass matrices for the SM fermions in the 4D effective theory
\begin{equation}\label{Eq:RSM}
(M^{RS}_f)_{ij} = \frac{v_W}{k r_c\pi}\lambda^f_{5,ij}
f^0_{L}(\pi,c^{L}_{f_i})f^0_{R}(\pi,c^{R}_{f_j}) \equiv
\frac{v_W}{k r_c\pi}\lambda^f_{5,ij}F_L(c^{L}_{f_i})F_R(c^{R}_{f_j}) \,,
\end{equation}
where $f$ labels the fermion species. The mass matrices are diagonalized by a bi-unitary transformation
\begin{equation}
(U_L^f)^\hc M^{RS}_f\,U_R^f =
\begin{pmatrix}
m^f_1 & 0         & 0 \\
0         & m^f_2 & 0 \\
0         & 0         & m^f_3
\end{pmatrix} \,,
\end{equation}
where $m^f_i$ are the masses of the SM quarks and leptons. The mass eigenbasis is then defined by
$\psi' = U^\hc\psi$, and the CKM matrix given by $V_{CKM} = (U^u_L)^\hc U^d_L$.

Because of KK interactions, the couplings of the $Z$ to fermions are shifted from their SM values. These shifts are not universal in general, which leads to flavor off-diagonal couplings when the fermions are rotated from the weak eigenbasis to the mass eigenbasis:
\begin{equation}
\mathcal{L}_\mrm{FCNC} \supset \frac{g_L}{c_W}Z_\mu\left\{
Q^L_Z(f')\sum_{a,b}\hat{\kappa}_{ab}^L\,\bar{f}'_{aL}\gamma^\mu f'_{bL}+
Q^R_Z(f')\sum_{a,b}\hat{\kappa}_{ab}^R\,\bar{f}'_{aR}\gamma^\mu f'_{bR}
\right\} \,.
\end{equation}
The mass eigenbasis is defined by $f' = U^\dag f$, and the flavour off-diagonal couplings are given by
\begin{equation}\label{Eq:kFCNC}
\hat{\kappa}_{ab}^{L,R} =
\sum_{i,j}(U^\dag_{L,R})_{ai}\kappa_{ij}^{L,R}(U_{L,R})_{jb} \,,
\end{equation}
where $\kappa_{ij}=\mrm{diag}(\kappa_1,\kappa_2,\kappa_3)$ are the coupling shifts due to KK interactions in the weak eigenbasis. Note that there would be no flavor violations if $\kappa$ is proportional to the identity matrix.

\begin{figure}[htbp]
\centering
\scalebox{1.15}[1]{\includegraphics[width=1.8in]{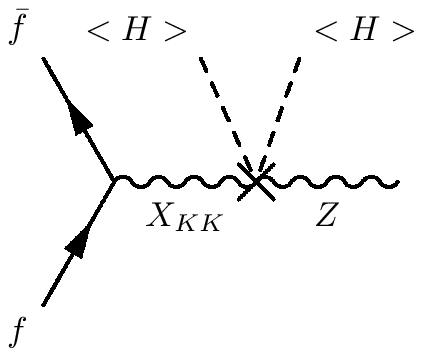}}
\hspace{0.5in}
\includegraphics[width=2.7in]{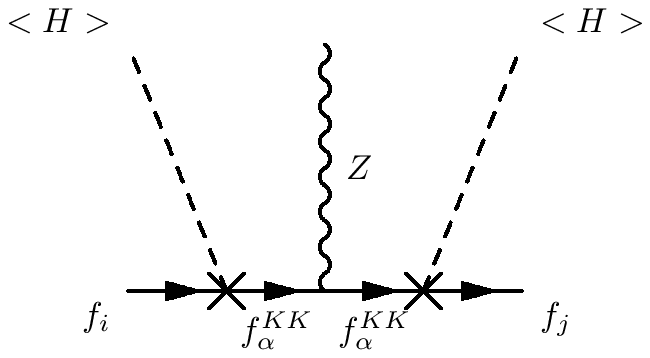}
\caption{\label{fig:KKZff} Effective $Z f\bar{f}$ coupling due to gauge and fermion KK mixings}
\end{figure}

The coupling shifts $\kappa_{ij}$ arises from two sources: mixing between the SM $Z$ and neutral KK gauge bosons, and that between the SM fermions and their KK excitations. These mixing processes are depicted in Fig.~\ref{fig:KKZff}.
In the gauge mixing diagram, $X$ can be either the SM $Z$ or the $Z'$ that arises from the $SU(2)_R$, while in the fermion mixing diagram, $i$, $j$, and $\alpha$ are generation indices. Note that the Yukawa interaction mixes LH (RH) SM fermions with the RH (LH) KK fermions. One can see from the diagrams in Fig.~\ref{fig:KKZff} that $\kappa_{ij}$ depend on the overlap of fermion wavefunctions, which are controlled by how the fermions are localized in the extra dimension. Details of the calculation and the full expressions of the contributions due to gauge and fermion KK mixings can be found in Ref.~\cite{CNW08}.

\section{\label{app:LepM} Typical admissible charged lepton mass matrices}
Parameterizing the complex 5D Yukawa couplings as $\lambda_{5,ij} = \rho_{ij}e^{i\phi_{ij}}$, admissible mass matrices of the forms given by Eq.~\eqref{Eq:RSM} are found with $\rho_{ij}$ and $\phi_{ij}$ randomly generated in the intervals $(0,2)$ and $[-\pi,\pi)$ respectively. In the following, for each of the five  charged lepton configurations determined by the c-parameter sets $c_L$ and $c_E$, we list for a typical solution that passes all current LFV constraints the mass eigenvalues, the magnitude of the complex mass matrix, $|M|$, of the solution, and the magnitude and phase of the corresponding 5D Yukawa coupling
matrix~\footnote{We need not list $\arg{M}$ because it is the same as $\phi$, as can be seen from Eq.~\eqref{Eq:RSM}.}. All numerical values are given to four significant figures (except for the mass eigenvalues, which required more accuracy to distinguish $m_\tau$). The mass eigenvalues agree with the charged masses at 1~TeV found in Ref.~\cite{XZZ08} to within one standard deviations quoted.
\begin{itemize}
\item Configuration~I:
\begin{gather}
c_L = \{0.6265, 0.5562, 0.5002\} \,, \quad
c_E = \{-0.6911,-0.5806,-0.5192\} \notag \\
\{m_e, m_\mu, m_\tau\} = \{4.95902 \cdot 10^{-4},0.104688,1.77976\}
\end{gather}
\begin{equation}
|M| =
\begin{pmatrix}
0.0003248 & 0.004970 & 0.04274 \\
0.001416  & 0.04607  & 0.3908  \\
0.009891  & 0.2738   & 1.717
\end{pmatrix}
\end{equation}
\begin{equation}
\rho =
\begin{pmatrix}
0.6598 & 0.2733 & 0.4445 \\
0.3272 & 0.2883 & 0.4624 \\
0.6022 & 0.4513 & 0.5351
\end{pmatrix} \,, \;
\phi =
\begin{pmatrix}
 0.2714 & -0.8414 & -3.055 \\
-1.509  &  2.910  & -1.787 \\
-1.294  &  1.547  &  0
\end{pmatrix}
\end{equation}

\item Configuration~II:
\begin{gather}
c_L = \{0.6474, 0.5802, 0.5172\} \,, \quad
c_E = \{-0.6714,-0.5566,-0.5026\} \notag \\
\{m_e, m_\mu, m_\tau\} = \{4.95902 \cdot 10^{-4},0.104688,1.78001\}
\end{gather}
\begin{equation}
|M| =
\begin{pmatrix}
0.0003856 & 0.008502 & 0.03564 \\
0.002473  & 0.04427  & 0.3681  \\
0.009663  & 0.2984   & 1.718
\end{pmatrix}
\end{equation}
\begin{equation}
\rho =
\begin{pmatrix}
0.7999 & 0.4591 & 0.5226 \\
0.5945 & 0.2770 & 0.6256 \\
0.4253 & 0.3419 & 0.5346
\end{pmatrix} \,, \;
\phi =
\begin{pmatrix}
-2.562 & -1.623  &  3.094   \\
-1.491 & -0.6484 & -0.05579 \\
-1.274 &  2.565  &  0
\end{pmatrix}
\end{equation}

\item Configuration~III:
\begin{gather}
c_L = \{0.6903, 0.5694, 0.5165\} \,, \quad
c_E = \{-0.6269,-0.5683,-0.5029\} \notag \\
\{m_e, m_\mu, m_\tau\} = \{4.95902 \cdot 10^{-4},0.104688,1.77966\}
\end{gather}
\begin{equation}
|M| =
\begin{pmatrix}
0.0006886 & 0.002653 & 0.009256 \\
0.01672   & 0.1594   & 0.5872   \\
0.03637   & 0.1366   & 1.670
\end{pmatrix}
\end{equation}
\begin{equation}
\rho =
\begin{pmatrix}
1.375  & 0.8461 & 0.5746 \\
0.6662 & 1.014  & 0.7273 \\
0.3606 & 0.2163 & 0.5147
\end{pmatrix} \,, \;
\phi =
\begin{pmatrix}
-2.730  & -2.449  &  0.7525 \\
-0.9022 & -0.2020 & -3.037  \\
 1.814  &  2.844  &  0
\end{pmatrix}
\end{equation}

\item Configuration~IV:
\begin{gather}
c_L = \{0.6908, 0.5786, 0.5142\} \,, \quad
c_E = \{-0.6277, -0.5583, -0.5057\} \notag \\
\{m_e, m_\mu, m_\tau\} = \{4.95902 \cdot 10^{-4},0.104688,1.77990\}
\end{gather}
\begin{equation}
|M| =
\begin{pmatrix}
0.0007459 & 0.002838 & 0.01386 \\
0.01879   & 0.1771   & 0.5452 \\
0.03762   & 0.2517   & 1.669
\end{pmatrix}
\end{equation}
\begin{equation}
\rho =
\begin{pmatrix}
1.556  & 0.6898 & 0.9268 \\
1.009  & 1.108  & 0.9381 \\
0.3642 & 0.2838 & 0.5176 
\end{pmatrix} \,, \;
\phi =
\begin{pmatrix}
 1.701 &  1.106 & -1.161 \\
-1.361 & -1.094 &  2.471 \\
 0     & -2.407 &  0.6682
\end{pmatrix}
\end{equation}

\item Configuration~V:
\begin{gather}
c_L = \{0.6477, 0.5689, 0.5002\}\,, \quad
c_E = \{-0.6714,-0.5681,-0.5190\} \notag \\
\{m_e, m_\mu, m_\tau\} = \{4.95902 \cdot 10^{-4},0.104688,1.78002\}
\end{gather}
\begin{equation}
|M| =
\begin{pmatrix}
0.0006072 & 0.005543 & 0.01604 \\
0.001143  & 0.03729  & 0.2536 \\
0.01666   & 0.4850   & 1.696
\end{pmatrix}
\end{equation}
\begin{equation}
\rho =
\begin{pmatrix}
1.268  & 0.4201 & 0.3330 \\
0.1955 & 0.2316 & 0.4316  \\
0.5200 & 0.5493 & 0.5265
\end{pmatrix} \,, \;
\phi =
\begin{pmatrix}
 0.5492  & -2.288 &  1.186  \\
 0.04108 & -1.505 & -1.032 \\
-1.560   &  2.667 & 0
\end{pmatrix}
\end{equation}
\end{itemize}

\end{document}